\documentclass[preprint,showpacs,preprintnumbers,amsmath,amssymb,aps]{revtex4}

\usepackage{graphicx}
\usepackage{dcolumn}
\usepackage{bm}
\usepackage{color}%

\begin{document}

\preprint{APS/123-QED}

\title{
Reconstruction of spatial qutrit states based on realistic measurement operators
}

\author{Gen Taguchi}\email{gentgch@hiroshima-u.ac.jp}
\author{Tatsuo Dougakiuchi}
\author{Masataka Iinuma}
\author{Holger F. Hofmann}
\author{Yutaka Kadoya}
\affiliation{%
Graduate School of Advanced Sciences of Matter, Hiroshima University, 
Higashihiroshima, 739-8530, Japan
}%

\date{\today}

\begin{abstract}
Spatial qudit states can be realized by using multi-slits to discretize the transverse momentum of a photon. 
The merit of this kind of spatial qudit states is that the implementation of higher dimensional  qudits is relatively easy. 
As we have recently shown, the quantum states of these spatial qudits can be analyzed by scanning a single interference pattern. 
This method of single scan tomography can also be applied at higher dimensions, but the reconstruction becomes more sensitive to smaller details of the scanned patterns as the dimensions increase. 
In this paper, we investigate the effect of finite measurement resolution on the single scan tomography of spatial qutrits. 
Realistic measurement operators describing the spatial resolution of the measurement are introduced and the corresponding pattern functions for quantum state reconstruction are derived. 
We use the pattern functions to analyze experimental results for entangled pairs of spatial qutrits generated by spontaneous parametric down-conversion (SPDC). 
It is shown that a reliable reconstruction of the quantum state can be achieved with finite measurement resolution if this limitation of the measurement is included in the pattern functions of single scan tomography. 
\end{abstract}


%
%
\maketitle
%
\section{Introduction}
In quantum information science, information processing tasks such as quantum cryptography, dense coding, and teleportation can not only be implemented using two-dimensional qubits, but are sometimes more efficiently performed using $d$-dimensional qudits as carriers of information \cite{BZZ2002,KB2002,KOCCEKO2003,DCGZ2003,WDLLL2005,HLB2002,OKBH2005,VWZ2002,LDHOPGBW2004,NLAMSP2005,TAZG2004,BLPK2005}. 
In optical implementations, we can make use of the degrees of freedom of photons to define the qubits or qudits. 
While photon polarization is a two-dimensional degree of freedom, the spatial (transverse momentum or angular momentum) and spectral (time-frequency) degrees of freedom are intrinsically continuous and can be used to define arbitrarily dimensional qudits by appropriate discretization. 
For example, qudits can be defined by selecting a set of $d$ angular momentum eigenstates represented by Gauss-Laguerre modes \cite{VWZ2002,LDHOPGBW2004}. 
Alternatively, multi-slits can be used to define $d$ transverse modes of the light field \cite{NLAMSP2005}. 
In this case, all modes are essentially equivalent and an increase of $d$ by adding more slits is relatively easy. 
Therefore, the method of using multi-slits to discretize the photons' transverse momentum is very promising for the realization of qudits with higher dimensions. 
\par
One of the challenges in realizing higher dimensional qudit systems is that the higher dimensional states require more precise and detailed measurements for analysis and control. 
For this purpose, it is useful to adapt the method of quantum state tomography to the specific qudit system \cite{LDHOPGBW2004,BLPK2005,WJEK1999,JKMW2001,TNWM2002,LTNDSP2007}. 
Recently, we have presented a method of quantum state tomography for $d$-slit spatial qudits based on the spatial patterns obtained by scanning the position of photon detection in a plane between the focal and image planes of a lens \cite{TDYKIHK2008}. 
In this initial work, we demonstrated the method for double-slit qubits and pointed out that the extension to higher dimensions is straightforward. 
However, as the following results for triple-slit qutrits show, an increase of the slit number $d$ requires an increased spatial resolution to observe the theoretically predicted patterns. 
Since such high spatial resolution is difficult to obtain, it is better to include the finite resolution of photon position detection in the method of quantum state tomography. 
It is then possible to obtain reliable results even though the measurement setup cannot fully resolve the details of the interference patterns. 
\par
In this paper, we present the formalism for single scan tomography with finite spatial resolution and demonstrate it experimentally for the case of spatial qutrits. 
For this purpose, we describe the measurements in terms of realistic measurement operators including the resolution of the detection setup. 
The elements of these measurement operators then provide the actual pattern functions observed in the experiments. 
The quantum state is reconstructed by fitting the set of pattern functions to the detection distribution. 
The effect of finite resolution is clearly seen in the improvement of the fit achieved by using the realistic measurement operators instead of the theoretical infinite resolution interference patterns. 
\par
The rest of the paper is organized as follows. 
In Sec. \ref{Th-A}, we review the principle of single scan tomography and show how finite resolution can be represented by the measurement operator. 
In Sec. \ref{Th-B}, we apply the method to two triple-slit qutrits measured by photon detection in a plane between the focal and image planes of a lens. 
We derive the pattern functions with and without finite resolution and illustrate the effects of finite resolution. 
In Sec. \ref{Ex1-A}, we describe our experimental setup and present measurement results that demonstrate the entanglement of the qutrit pairs. 
In Sec. \ref{Ex1-B}, we present and discuss the experimental results for single scan tomographies of spatial qutrit states prepared by conditional measurements on the other qutrit. 
The complete density matrix of the two qutrits is then reconstructed from a sufficiently large set of conditional measurements. 
Sec. \ref{Conclusion} concludes the paper. 
%
%
\section{Single Scan Tomography with Finite Spatial Resolution\label{Th-A}}
Multi-dimensional spatial qudits can be defined by discretizing photon transverse momentum using multi-slits. 
If the quantum state of the photon passing through the $i$-th slit is given by $|s_i\rangle$, a $d$-slit generates a qudit described by the orthogonal basis $\{|s_i\rangle\}$. 
After the state of the photon is discretized by the slits, the photon propagates in continuous free space. 
The state of the spatial qudit can then be analyzed by measurements of photon position in this continuous space. 
Specifically, the probability of detection at a point $x$ in any plane behind the $d$-slit is given by a measurement operator $\hat{M}(x)$ that acts on the $d$-dimensional Hilbert space of the spatial qudit. 
For a qudit state given by a density matrix $\hat{\rho}$, the probability of detection is 
\begin{eqnarray}
P(x)\!&=&\!
\textrm{Tr}\left[\hat{M}(x)\hat{\rho}\right]
=
\sum_{i,j}{M}_{ij}(x)\rho_{ji}
\;, 
\label{TraceRhoM}
\end{eqnarray}
where ${M}_{ij}(x)$ and $\rho_{ji}$ are the matrix elements of the operators in the basis of the spatial qudit $\{|s_i\rangle\}$. 
Thus the detection probability $P(x)$ is a linear combination of the measurement operator elements with the coefficients given by the density matrix elements of the qudit state. 
If all of the pattern functions ${M}_{ij}(x)$ are linearly independent, it is possible to invert and solve the equation for the density matrix elements, so that the complete density matrix can be determined from the scanned distribution $P(x)$. 
This is the principle of single scan tomography. 
\par
The measurement operator can be derived by projecting the operator $\hat{\Pi}(x)$ describing the position measurement with outcome $x$ into the Hilbert space of the spatial qudit. 
\begin{eqnarray}
\hat{M}(x) \!&=&\! \sum_{i,j}|{s_i}\rangle\langle{s_i}|\hat{\Pi}(x)
|{s_j}\rangle\langle{s_j}|
 = 
\sum_{i,j}{M}_{ij}(x)|{s_i}\rangle\langle{s_j}|
. 
\label{Eq:MeasurementOperatorExpression}
\end{eqnarray}
Since $x$ is a continuous variable, all realistic measurements will have a finite resolution. 
This resolution can be described by the measurement operator $\hat{\Pi}(x)$. 
In the present experiment, the position measurement is realized by a slit of width $b$, so that $\hat{\Pi}(x)$ is given by the integral 
\begin{eqnarray}
\hat{\Pi}(x)=\frac{1}{b}\int_{-b/2}^{b/2}|{x+x'}\rangle\langle{x+x'}|dx'
.
\label{Eq:PositionMeasurementOperator}
\end{eqnarray}
If $\varphi_i(x) = \langle{x}|{s_i}\rangle$ is the transverse wave function originating from the $i$-th slit, ${M}_{ij}(x)$ can be expressed by
\begin{eqnarray}
{M}_{ij}(x)=\frac{1}{b}\int_{-b/2}^{b/2}\varphi_i^*(x+x')\varphi_j(x+x')dx
. 
\label{Eq:POVM_Element}
\end{eqnarray}
In the ideal case with infinite resolution, $\hat{\Pi}(x)$ is the projection operator $|{x}\rangle\langle{x}|$ and the idealized measurement operator element is given by
\begin{eqnarray}
\lim_{b \to 0}{M}_{ij}(x) \!&=&\! \varphi_i^*(x)\varphi_j(x)
. 
\label{Eq:Ideal_POVM_Element}
\end{eqnarray}
These idealized patterns have the highest possible visibility. 
The finite resolution reduces the visibility of the measurement. 
However, quantum state reconstruction only depends on the linear independence of the patterns. 
Therefore the realistic pattern functions of Eq. (\ref{Eq:POVM_Element}) result in a more precise reconstruction of the density matrix than the idealized pattern functions of Eq. (\ref{Eq:Ideal_POVM_Element}). 
%
\section{Measurement in the Intermediate Plane\label{Th-B}}
\begin{figure}
\begin{center}
\includegraphics[width=0.720\linewidth,height=0.299\linewidth]{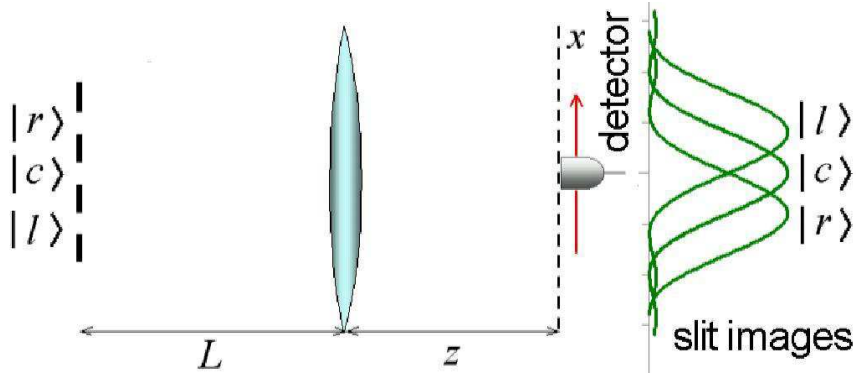}
\caption{\label{fig:SchematicSetup} 
Measurement between the focal and image planes. 
}
\end{center}
\end{figure}

The measurement of the spatial qudit is performed by detection in the intermediate plane between the focal and image planes of a lens. 
The lens of focal length $f$ is placed at a distance of $L$ from the slit and the photon is detected in the plane at a distance $z$ from the lens, where $z$ is between the focal plane and the image plane of the lens. 
Figure \ref{fig:SchematicSetup} shows the schematic setup in the case of qutrits, $d=3$, with the slit basis given by $\{|l\rangle, |c\rangle, |r\rangle\}$, each element of which corresponds to the photon passing through the left, center, or right slit, respectively. 
%
%
\par
The wave function $\varphi_i(z;x)$ can be calculated using the Fresnel-Kirchhoff diffraction integral for an effective distance of $R = (Lf+zf-Lz)/(z-f)$ from the multi-slit \cite{TDYKIHK2008}. 
If the detection plane is far enough from the image plane, so that the slit width $a$ is sufficiently smaller than $\sqrt{R\lambda}$, where $\lambda$ is the wave length of the photon, the wave function takes the form of the conventional slit diffraction pattern given by the sinc function $\textrm{sinc}(x)=\sin x/x$, 
\begin{eqnarray}
\varphi_i(z;x)
\!&=&\!\sqrt{\frac{K}{\pi}}
e^{-i\frac{2 r_i}{a}K x}
\textrm{sinc}\left[K\left(x\!+\!\frac{z-f}{f}r_i\right)\right]
. 
\label{ApproxWavFunc}
\end{eqnarray}
Here, $K = \pi a f/\lambda R(z-f)$ scales the sinc function and $r_i$ is the distance between the optical axis and the $i$-th slit \cite{TDYKIHK2008}. 
According to Eq. (\ref{TraceRhoM}), the detection probability is 
\begin{eqnarray}
P(x)
\!&=&\!
\rho_{ll}M_{ll}(x)
\!\!+2\textrm{Re}\left[\rho_{lc}\right]\textrm{Re}\left[M_{lc}(x)\right]
+2\textrm{Im}\left[\rho_{lc}\right]\textrm{Im}\left[M_{lc}(x)\right]
\nonumber\\\!&{}&\!
\!\!+2\textrm{Re}\left[\rho_{lr}\right]\textrm{Re}\left[M_{lr}(x)\right]
+2\textrm{Im}\left[\rho_{lr}\right]\textrm{Im}\left[M_{lr}(x)\right]
\!+\rho_{cc}M_{cc}(x)
\nonumber\\\!&{}&\!
\!\!+2\textrm{Re}\left[\rho_{cr}\right]\textrm{Re}\left[M_{cr}(x)\right]
+2\textrm{Im}\left[\rho_{cr}\right]\textrm{Im}\left[M_{cr}(x)\right]
\!+\rho_{rr}M_{rr}(x)
. 
\label{TraceRhoM_Qutrit}
\end{eqnarray}
In the ideal infinite resolution case for a spatial qutrit, the pattern functions are given by Eq. (\ref{Eq:Ideal_POVM_Element}). 
In the realistic case of finite resolution, the pattern functions are given by the integral in Eq. (\ref{Eq:POVM_Element}) that describes the reduced visibility of the interference patterns. 
\par
\begin{figure*}
\begin{center}
\includegraphics[width=1.00\linewidth,height=0.812\linewidth]{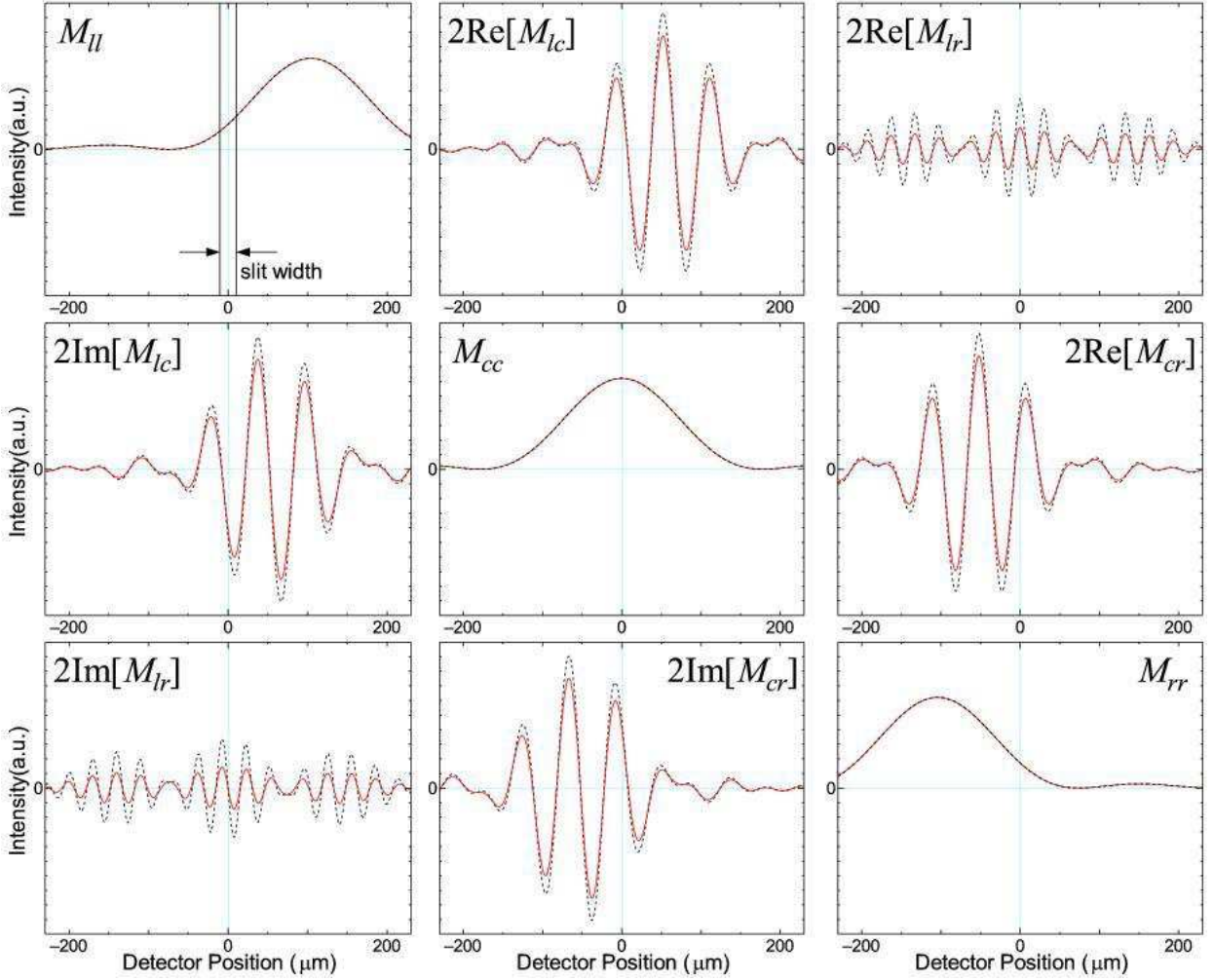}
\caption{\label{fig:PatternFunctions} 
The nine pattern functions $M_{ij}(x)$ for a spatial qutrit. 
Dashed lines correspond to ideal measurements with infinite resolution. 
Solid lines correspond to finite resolution measurements. 
The slit width of the detector system indicated in the upper left hand graph defines the resolution of the realistic measurements. 
}
\end{center}
\end{figure*}

To illustrate the effects of finite resolution in the detection of the photon position, Fig. \ref{fig:PatternFunctions} shows the pattern functions for infinite and finite resolution of the detector system. 
The dashed lines correspond to ideal measurements with infinite resolution and the solid lines correspond to finite resolution measurements. 
The parameters in the figure were chosen to correspond to the experiments described in Section \ref{Ex1-A}. 
The detector slit width of 20 $\mu$m that defines the resolution is indicated in the first of the nine graphs. 
The main effect of the finite resolution is to reduce the intensity of the oscillating pattern functions. 
The higher the frequency of oscillation, the more the intensity is reduced. 
Since the diagonal element patterns $M_{ii}$ do not oscillate, their change is negligible. 
In general, the visibilities of interference between slits are weakened more if the slits are farther apart. 
However, the determination of the correct density matrix elements only depends on the use of the correct pattern functions, not on their visibilities. 
%
%
%
\section{Experimental Setup and Verification of Entanglement\label{Ex1-A}}
In our experiments, we demonstrate the spatial qutrit analysis using entangled photon pairs generated by SPDC. 
A lens images the transverse momenta of the down-converted photons onto a pair of triple-slits. 
After the triple-slits, the photon pairs should ideally be in the maximally entangled qutrits states given by
\begin{eqnarray}
|\Psi\rangle_{AB}
\!&=&\!
\frac{1}{\sqrt{3}}
\big(|l\rangle_{A}|r\rangle_B+|c\rangle_A|c\rangle_B
+|r\rangle_A|l\rangle_B\big)
, 
\label{EntangledState}
\end{eqnarray}
where the suffixes A and B denote the photons in the different paths. 
\par
\begin{figure*}
\begin{center}
\includegraphics[width=0.924\linewidth,height=0.469\linewidth]{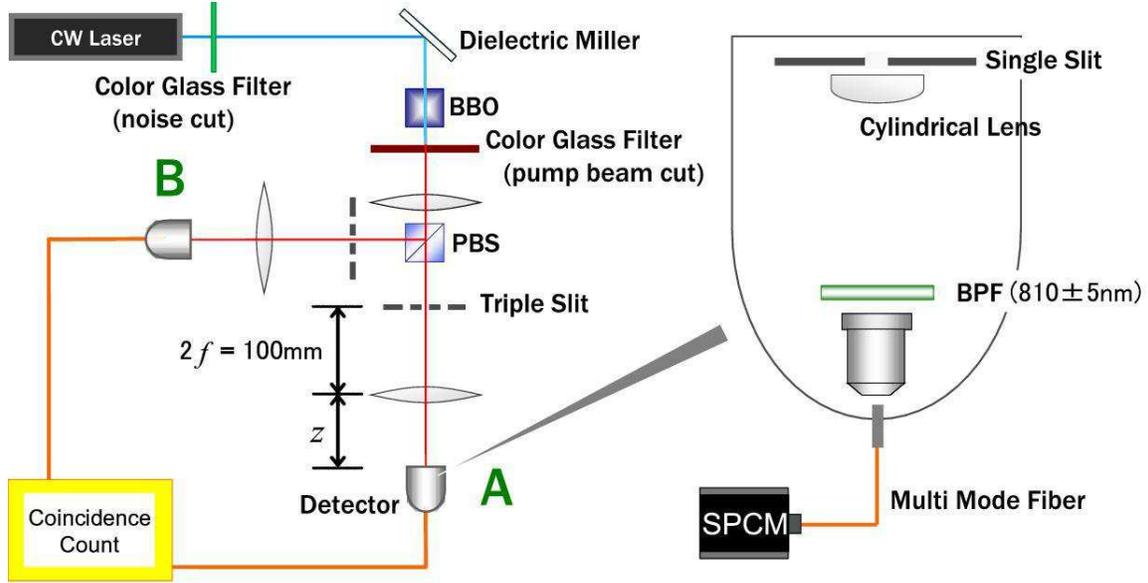}
\caption{\label{fig:setup}
Experimental setup. 
The transverse momentum is selected by the triple slit in the focal plane of the first lens. 
Photons are separated into two arms by the PBS. 
The second lens is placed at a distance of 2$f$ from the triple slit in each arm. 
Coincidence measurements are performed using the detector system shown on the right. 
}
\end{center}
\end{figure*}

The experimental setup is shown in Fig. \ref{fig:setup}. 
The entangled photon pairs were generated by pump beam from a 45 mW cw laser with 405 nm wavelength incident on a 5mm-thick $\beta$-barium borate (BBO) crystal. 
The BBO cristal was set for Type II SPDC in the collinear condition. 
The photons were separated into two different arms by a polarizing beam splitter (PBS). 
The lens before the PBS focuses the momentum eigenstates of the photons on a pair of triple-slits, one in each path. 
The slit width was 45 $\mu$m and the distance between the slits was 135 $\mu$m. 
To measure the spatial qutrits generated by the triple-slits, a lens of focal length $f$ = 50 mm was placed at a distance of $L=2f$ from each of the triple slits. 
A detector system was placed at a distance of $z$ from each of the lenses. 
The detector system was constructed as shown in inset of Fig. \ref{fig:setup}. 
Single slits of width 20 or 40 $\mu$m were used to select the transverse positions of the photons and an objective lens behind a band pass filter of band width 810$\pm$5 nm coupled the photons into the multimode fiber connected to a photon detector (Perkin-Elmer SPCM-AQR-14) recording the coincidence counts between the two arms. 
Cylindrical lenses were used to reduce the coupling loss between the single slit and the objective lens caused by diffraction at the 20 $\mu$m single slits. 
\par
\begin{figure*}
\begin{center}
\includegraphics[width=68.4mm,height=60.04mm]{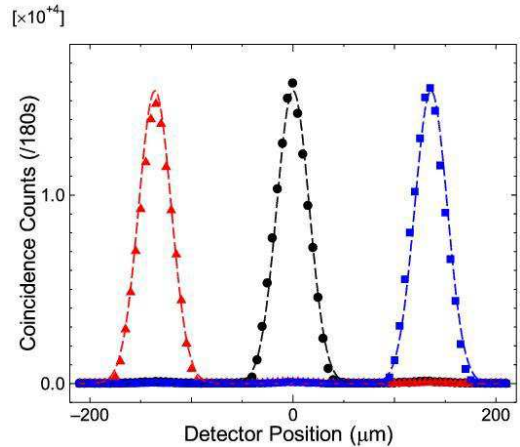}
\caption{\label{Image_plane_measurement}
Coincidence count rates as a function of scanning detector position for which-path correlations measured in the image plane. 
The triangles, dots, and squares show the experimental data corresponding to the fixed detector positions at the left, center, and right slit image in arm B, respectively. 
The dashed lines are guides for the eyes. 
}
\end{center}
\end{figure*}
\begin{figure*}
\begin{center}
\includegraphics[width=140.07mm,height=55.5mm]{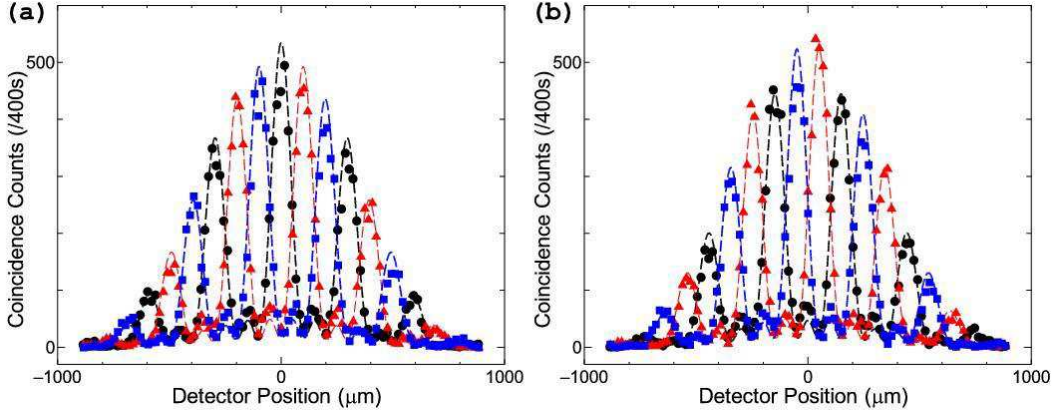}
\caption{\label{Focal_plane_measurement}
Coincidence count rates as a function of scanning detector position for interference correlations measured in the focal plane. 
The dots, squares, and triangles show the experimental data and the dashed lines are guides for the eyes. 
(a) The detector in arm B was set to the positions corresponding to phase differences of $-{\pi / 3}$ (triangles), 0 (dots), and ${\pi / 3}$ (squares). 
(b) The detector in arm B was set to the positions corresponding to phase differences of $-{\pi / 6}$ (triangles), ${\pi / 6}$ (squares), and ${\pi / 2}$ (dots).}
\end{center}
\end{figure*}

To verify whether our photon source produces the intended entanglement given by Eq. (\ref{EntangledState}), we first confirmed the which-path and interference correlations in the image and focal planes of the lenses between the triple-slits and the detector systems. 
Since the resolution was not critical for this purpose, a detector slit width of 40 $\mu$m was used in these experiments. 
The which-path correlation was observed in image plane measurements. 
The detector position $x$ in arm A was scanned, while the detector in arm B was fixed at the center of one of the slit images. 
The results are shown in Fig. \ref{Image_plane_measurement}. 
In each scan, the peak appears only at the position corresponding to the fixed detector position in arm B, confirming the expected correlations between the spatial qutrits. 
The interference correlation was observed in focal plane measurement. 
In this case, the position in arm B corresponds to a superposition of all three slits with a specific phase relation selected by the choice of position. 
The interference patterns obtained are shown in Fig. \ref{Focal_plane_measurement}. 
The phases of the interference patterns clearly depend on the detector position in arm B, again confirming the expected correlations between the spatial qutrits. 
These two measurements verify that the state generated by our photon source in is indeed close to the entangled state given by Eq. (\ref{EntangledState}). 
\par
%
%
%
%
%
\section{Density Matrix Reconstruction by Single Scan Tomography \label{Ex1-B}}
To obtain the complete density matrix of the entangled qutrits, we now apply single scan tomography to our photon source. 
For this purpose, we have to select an intermediate plane between the focal and image planes. 
When the detection plane is near the image plane, which-path information corresponding to diagonal elements of the density matrix is observed most clearly. 
Interference information corresponding to off-diagonal elements is easier resolved in measurements near the focal plane. 
In principle, there should be an optimal distance $z$ for single scan tomography. 
Based on our previous work \cite{TDYKIHK2008}, we started from a distance of $z = 1.8f$, but found that some improvements in the results of the tomography could be achieved by moving the detectors to a distance of $z = 1.81f$ from the lens. 
The pattern functions shown in Fig. \ref{fig:PatternFunctions} of Sec. \ref{Th-B} were calculated using this distance between lens and detector. 
\par
As the discussion in Sec. \ref{Th-B}, single scan tomography is sensitive to the resolution of position detection. 
To achieve as high a resolution as possible, the single slit in the detector system was changed to 20 $\mu$m and the cylindrical lens was placed behind the slit to compensate the diffraction effects. 
However, there remains a non-negligible loss of visibility due to the finite slit width, as shown by the pattern functions in Fig. \ref{fig:PatternFunctions}. 
\par
We have investigated the effects of finite resolution by performing single scan tomography of conditional denisity matrices in arm A generated by measurements at fixed detector positions in arm B. 
A comparison of single scan tomography using idealized pattern functions neglecting the slit width with tomography using realistic pattern functions including the slit width is shown in Fig. \ref{fig:ConditinalScanningData0}. 
Here, the fixed detector in arm B was near the optical axis at 0 $\mu$m
\footnote{The actual center of the interference pattern in arm B was found at $-1.1 \mu$m according to a least square fit to the experimental data.}. 
Figure \ref{fig:ConditinalScanningData0} (a) shows the coincidence counts obtained by scanning the detector position in arm A. 
%
\unitlength=0.001\linewidth
\begin{figure}
\begin{picture}(840,340)
\put(0,0){\makebox(840,340){\includegraphics{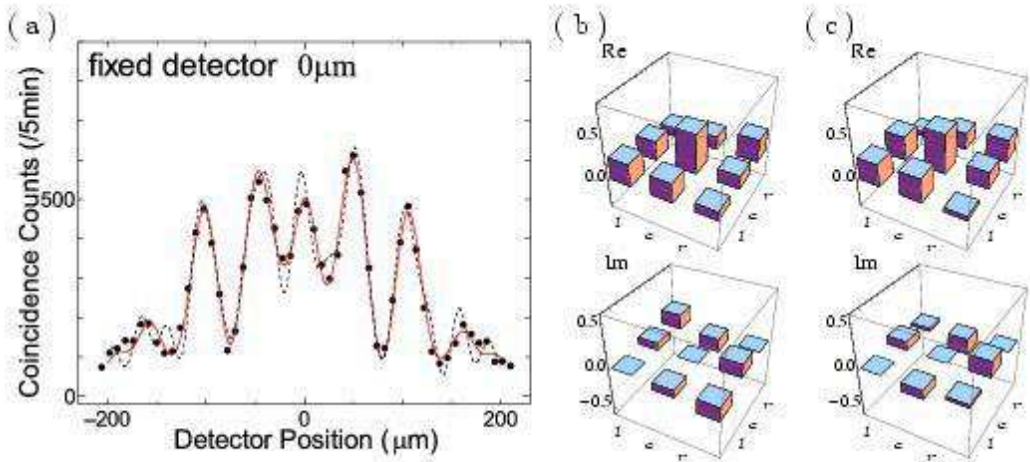}}}
\end{picture}
\caption{\label{fig:ConditinalScanningData0} 
Comparison of single scan tomography with and without slit width. 
(a) Coincidence counts obtained in arm A when the fixed detector is near the optical axis in arm B. 
The dashed line shows the fit using the idealized pattern functions for negligible slit width. 
The solid line shows the fit using the realistic pattern functions for the slit width of 20 $\mu$m used. 
(b) Density matrix reconstructed with idealized pattern functions. 
(c) Density matrix reconstructed with realistic pattern functions. 
}
\end{figure}
%
We performed two density matrix reconstructions, one using the idealized patterns for negligible slit width $b{\to}0$ defined by Eq. (\ref{Eq:Ideal_POVM_Element}) and the other using the realistic pattern functions defined by Eq. (\ref{Eq:POVM_Element}). 
The dashed line shows the fit with the idealized pattern functions and the solid line shows the fit with the realistic pattern functions. 
Interestingly, the fit to the details of the experimental data improves significantly when the finite resolution is taken into account. 
Thus, the application of realistic pattern functions is not only a matter of quantitative errors, but also influences the correct identification of characteristic features with elements of the density matrix. 
This effect is illustrated by the comparison of the density matrices reconstructed from idealized pattern functions shown in Fig. \ref{fig:ConditinalScanningData0}(b) with the more precise density matrix reconstructed from realistic pattern functions shown Fig. \ref{fig:ConditinalScanningData0}(c). 
The density matrices differ in the values obtained for the off-diagonal elements. 
In the element $\rho_{lr}$, the change not only affects the size but also the phase of the reconstructed value. 
Thus, the patterns $M_{lr}(x)$ have a rathere subtle effect on the shapes of the reconstructed density matrices. 
\par
We performed a sufficient number of conditional density matrix reconstructions for different setting in arm B to reconstruct the complete $9\times9$ density matrix of the composite system, $\hat{\rho}_{AB}$. 
Figure \ref{fig:ConditinalScanningData} shows a selection of the conditional scans obtained at different detector positions in arm B, together with the density matrices reconstructed using realistic pattern functions. 
These scans provide an experimental illustration of the relation between the scanned distribution and the density matrix elements. 
In total, we performed 29 single scan tomographies with different detector positions in arm B. 
We could then determine the precise values of experimental parameters such as $L$ by least square fits. 
The reconstructed density matrix of the complete entangled two qutrits system was found to be 
\tiny
\begin{eqnarray}
&{}&\hspace{-5mm}
\hat{\rho}_{AB}
\;=\;
\nonumber
\\
&{}&\hspace{-5mm}
\frac{1}{100}
\left(\begin{array}{c c c c c c c c c}
 \!\;\;\;1.7 \!\!&\!\! -0.1-0.5i \!\!&\!\!  \;\;\;1.0+0.4i \!\!&\!\!  \;\;\;0.5-0.5i \!\!&\!\!  \;\;\;0.3-0.4i \!\!&\!\!  \;\;\;0.8-0.7i \!\!&\!\! -0.4+0.0i \!\!&\!\! -1.3+0.7i \!\!&\!\!  \;\;\;3.2-0.2i\\
 \!-0.1+0.5i \!\!&\!\!  \;\;\;4.8 \!\!&\!\! -0.5+0.8i \!\!&\!\!  \;\;\;1.0-0.6i \!\!&\!\!  \;\;\;0.7+1.0i \!\!&\!\!  \;\;\;0.1+0.3i \!\!&\!\! -1.0+0.0i \!\!&\!\!  \;\;\;1.1-0.7i \!\!&\!\!  \;\;\;0.1-0.1i\\
 \!\;\;\;1.0-0.4i \!\!&\!\! -0.5-0.8i \!\!&\!\!  \colorbox[gray]{0.95}{28.1} \!\!&\!\!  \;\;\;0.2-0.2i \!\!&\!\!  \colorbox[gray]{0.95}{$\;\;\;29.6-5.8i$} \!\!&\!\! -0.3+0.2i \!\!&\!\!  \colorbox[gray]{0.95}{$\;\;\;24.7-4.7i$} \!\!&\!\! -0.5+0.8i \!\!&\!\! -0.4-0.1i\\
 \!\;\;\;0.5+0.5i \!\!&\!\!  \;\;\;1.0+0.6i \!\!&\!\!  \;\;\;0.2+0.2i \!\!&\!\!  4.7 \!\!&\!\!  \;\;\;0.3+1.0i \!\!&\!\! -1.4-0.7i \!\!&\!\! -0.4+0.0i \!\!&\!\! -0.5-0.0i \!\!&\!\!  \;\;\;0.4-1.0i\\
 \!\;\;\;0.3+0.4i \!\!&\!\!  \;\;\;0.7-1.0i \!\!&\!\!  \colorbox[gray]{0.95}{$\;\;\;29.6+5.8i$} \!\!&\!\!  \;\;\;0.3-1.0i \!\!&\!\!  \colorbox[gray]{0.95}{22.3} \!\!&\!\!  \;\;\;0.6-0.4i \!\!&\!\!  \colorbox[gray]{0.95}{$\;\;\;29.7+0.7i$} \!\!&\!\!  \;\;\;0.7-0.3i \!\!&\!\!  \;\;\;0.2+0.4i\\
 \!\;\;\;0.8+0.7i \!\!&\!\!  \;\;\;0.1-0.3i \!\!&\!\! -0.3-0.2i \!\!&\!\! -1.4+0.7i \!\!&\!\!  \;\;\;0.6+0.4i \!\!&\!\!  4.1 \!\!&\!\! -0.3+1.5i \!\!&\!\! -2.4-0.7i \!\!&\!\!  \;\;\;0.3+0.3i\\
\!-0.4-0.0i \!\!&\!\! -1.0+0.0i \!\!&\!\!  \colorbox[gray]{0.95}{$\;\;\;24.7+4.7i$} \!\!&\!\! -0.4-0.0i \!\!&\!\!  \colorbox[gray]{0.95}{$\;\;\;29.7-0.7i$} \!\!&\!\! -0.3-1.5i \!\!&\!\!  \colorbox[gray]{0.95}{27.4} \!\!&\!\!  \;\;\;0.1+0.8i \!\!&\!\!  \;\;\;0.6+0.9i\\
\!-1.3-0.7i \!\!&\!\!  \;\;\;1.1+0.7i \!\!&\!\! -0.5-0.8i \!\!&\!\! -0.5+0.0i \!\!&\!\!  \;\;\;0.7+0.3i \!\!&\!\! -2.4+0.7i \!\!&\!\!  \;\;\;0.1-0.8i \!\!&\!\!  5.0 \!\!&\!\!  \;\;\;0.4+0.4i\\
 \!\;\;\;3.2+0.2i \!\!&\!\!  \;\;\;0.1+0.1i \!\!&\!\! -0.4+0.1i \!\!&\!\!  \;\;\;0.4+1.0i \!\!&\!\!  \;\;\;0.2-0.4i \!\!&\!\!  \;\;\;0.3-0.3i \!\!&\!\!  \;\;\;0.6-0.9i \!\!&\!\!  \;\;\;0.4-0.4i \!\!&\!\!  1.9
\end{array}\right)
. 
\nonumber
\label{NumericalDensityMatrixS}
\end{eqnarray}
\normalsize
Figure \ref{fig:DensityMatrix1} shows a comparison of this density matrix with the ideal pure state density matrix of the maximally entangled state given by Eq. (\ref{EntangledState}). 
The fidelity of the ideal state $|\Psi\rangle_{AB}$ was $F={}_{AB}\langle \Psi|\hat{\rho}_{AB}|\Psi\rangle_{AB}=0.819$. 
Thus the inclusion of position resolution defined by the slit width in the pattern functions permits a reliable reconstruction of the quantum state in the nine dimensional Hilbert space of the entangled qutrit pair. 
\unitlength=0.001\linewidth
\begin{figure}
\begin{picture}(1000,900)
\put(0,0){\makebox(1000,900){\includegraphics{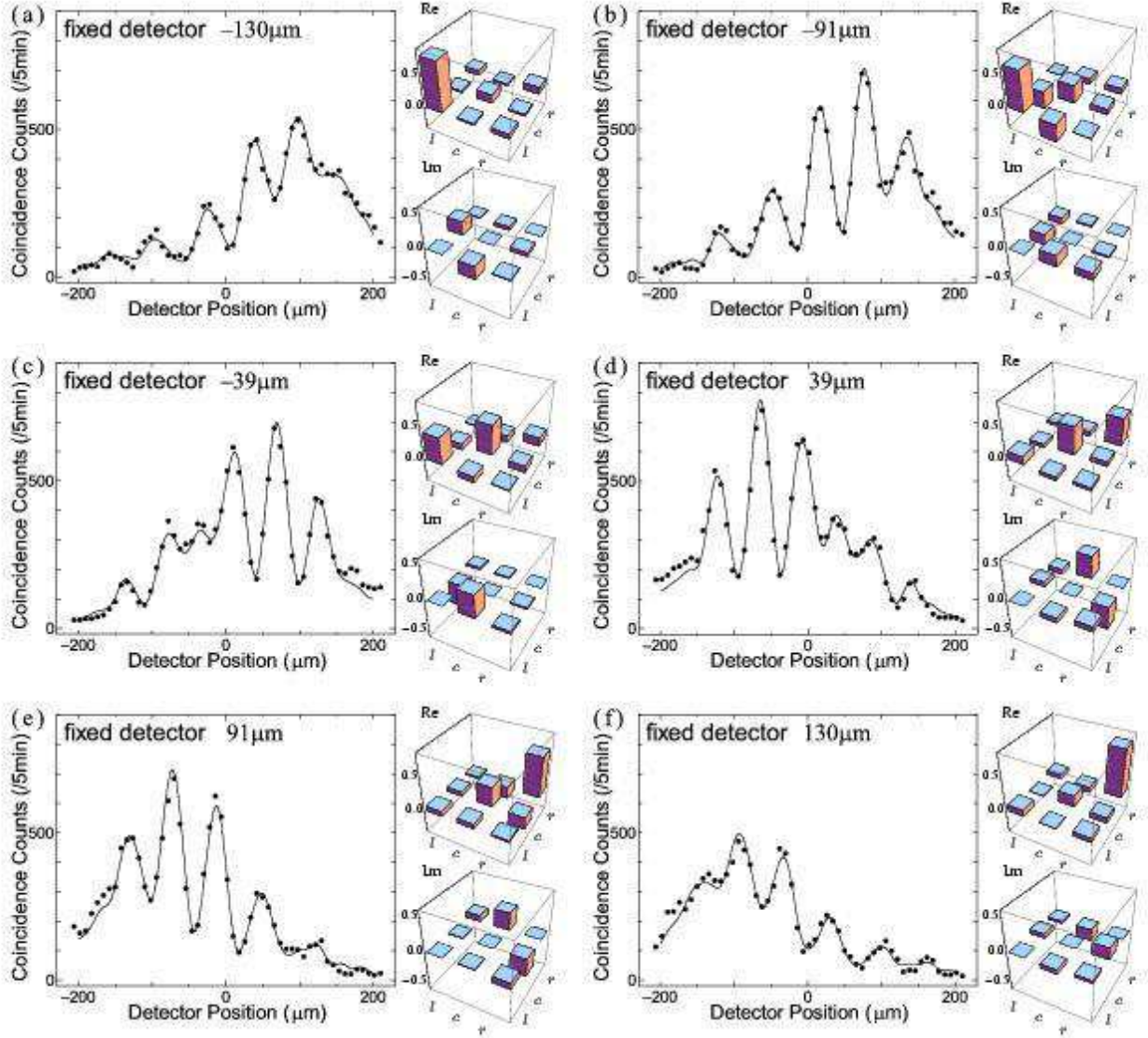}}}
\end{picture}
\caption{\label{fig:ConditinalScanningData} 
Coincidence counts data. 
Each measurement was obtained by placing the fixed detector in a different position. 
The solid lines are fitted to the data points using realistic pattern functions icluding the effects of finite slit width. 
The corresponding conditional density matrix is shown on the right side each interference pattern. 
}
\end{figure}
\unitlength=0.001\linewidth
\begin{figure}
\begin{picture}(640,640)
\put(0,0){\makebox(640,640){\includegraphics{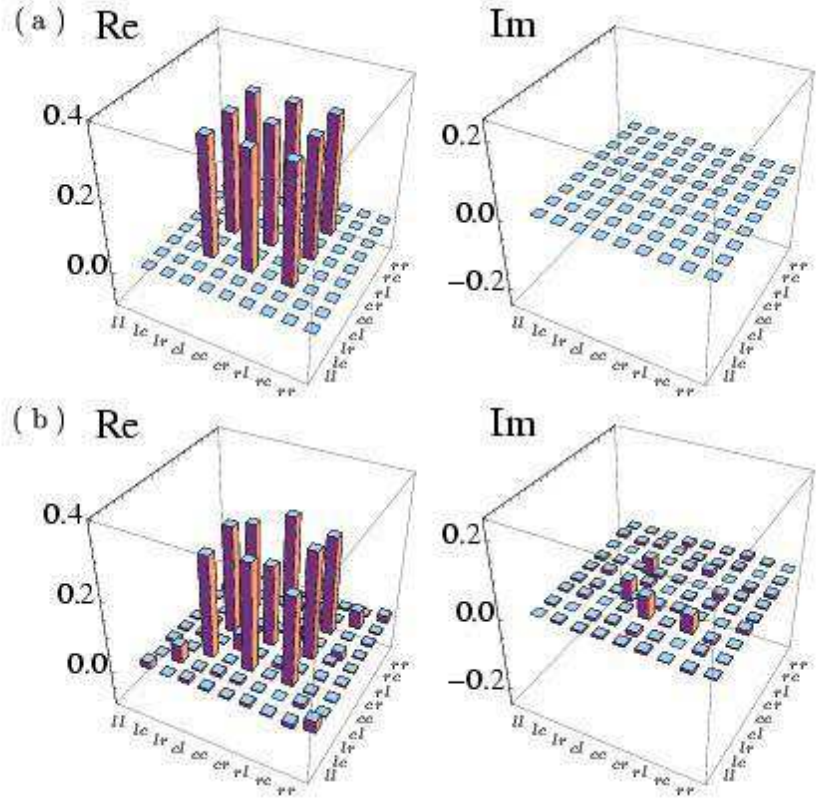}}}
\end{picture}
\caption{ 
Real and imaginary part of 
(a) the density matrix of the maximally entangled pure state and 
(b) the density matrix reconstructed from single scan tomographies with realistic pattern functions. 
}
\label{fig:DensityMatrix1}
\end{figure}
%
%
\section{Conclusions\label{Conclusion}}
Single scan tomography is a method of state reconstruction for an arbitrary dimensional spatial qudit that directly connects the scanned distribution to the density matrix elements. 
We can visually understand the elements of the measurement operators as pattern functions and a fit of the scanned distribution using these pattern functions provides the reconstructed density matrix elements. 
A reliable reconstruction is possible whenever the pattern functions associated with different density matrix elements are linearly independent. 
It is therefore possible to include the effects of finite measurement resolution in the pattern functions without affecting the reconstruction of the density matrix. 
The results obtained with realistic pattern functions based on measurement operators describing the finite resolution of the experiment are therefore better than the ones obtained with idealized pattern functions based on infinitely precise measurement. 
\par
We have demonstrated the effect of including the finite resolution in the pattern function on the experimental reconstruction of spatial qutrits. 
The results show that a reliable reconstruction of the density matrix is possible even though the finite spatial resolution of 20 $\mu$m has a non-negligible effect on the pattern functions. 
In particular, a fidelity of 0.819 could be obtained for the entangled qutrit pair when the realistic pattern functions were used in the reconstruction of the density matrix. 
\par
As the dimensionality of spatial qudit systems increases, it will be increasingly difficult to realize high resolution measurements. 
The method of single scan tomography with finite measurement resolution may therefore play a significant role in the realization of higher dimensional qudit systems. 
\begin{acknowledgments}
We are grateful to Hiroshi Taniguchi for his technical support. 
Part of this work was supported by the Grant-in-Aid program of the Japanese Society for the Promotion of Science. 
\end{acknowledgments}
\newpage 

\end{document}